\documentclass[11pt]{article}

\usepackage{geometry}                
\usepackage{adjustbox}
\usepackage{booktabs}
\usepackage[table]{xcolor}
\usepackage{siunitx}
\usepackage{amsmath}
\usepackage{array}
\usepackage{graphicx}
\usepackage{amssymb}
\usepackage{epstopdf}
\usepackage{float}
\usepackage[english]{babel}
\usepackage[latin1]{inputenc}
\usepackage{filecontents}
\usepackage{caption}
\usepackage{authblk}

\begin{document}

\title{Modeling sexual selection in T\'ungara frog and rationality of mate choice}

\date{}
\author[1]{Esteban Vargas}
\author[1]{Camilo Sanabria}

\affil[1]{Department of Mathematics, Universidad de los Andes, Bogota, Colombia}

\maketitle

\begin{abstract}
The males of the specie of frogs \textit{Engystomops pustulosus} produce simple and complex calls to lure females, as a way of \textit{Intersexual selection}. Complex calls lead males to a greater reproductive success than simple calls do. However, the complex calls are also more attractive to the main predator of these amphibians, the bat \emph{Trachops cirrhosus}. Therefore, as M. Ryan suggests in \cite{ryan1985tungara}, the complexity of the calls let the frogs keep a trade off between reproductive success and predation. In this paper, we first propose to model the proportion of simple to complex calls as a symmetric game of two strategies. We also propose a model with three strategies (simple callers, complex callers and quiet males), where we assess the effect of a male that keeps quiet and intercepts females, which would play a role of \textit{Intrasexual selection}. We analyze the stable points of the replicator equations of the models that we propose. Under the assumption that the decision of the males takes into account this trade off between reproductive success and predation, our model reproduces the observed behavior reported in the literature with minimal assumption on the parameters. From the three strategies model, we verify that the quiet strategy could only coexists with the simple and complex strategies as long as the rate at which quiet males intercept females is high. We conclude that the reproductive strategy of the male frog \textit{Engystomops pustulosus} is rational.
\end{abstract}


\section{Introduction}

T\'ungara frog (\textit{Engystomops pustulosus}) is an example of a specie where males have developed displays in order to attract females more efficiently. In consequence, these amphibians are an example of Sexual Selection, as Darwin referred. T\'ungara males display a great variety of calls. As females chooses the males base on the complexity of the call (as M. Ryan demonstrates in \cite{ryan1985tungara}), we can refer to this kind of selection as \textit{Intersexual selection} (\cite{andersson1994sexual}). The simplest one is a low frequency sound which lasts approximately $400$ $ms$ and it is called \textit{whine}. We refer to this call as \textit{simple call}. Males add higher frequency sounds called \textit{chucks} to the whine call in order to create more complex calls. Each chuck lasts around $42$ $ms$ and one male can add from one to seven chucks to the whine to form what we refer as a \textit{complex call}. All males are equally capable of producing the same calls, so bias in the proportion of a certain call is due to males decision (\cite{ryan1985tungara}). Most males form choruses along the shore of the ponds. The choruses comprise call bouts and pauses in between. Based on the studies of X.E. Bernal \textit{et al.} \cite{bernal2007cues}, in those choruses, t\'ungara frogs produce over 167 calls in three minutes, from which 70 percent are simple calls and the remaining are complex calls. The bouts are started by focal males. A focal male produces over 26 calls per call bout, which last over 66 seconds. From those calls, 53 percent are simple. In \cite{akre2011female} it is shown that when females are close to the choruses, the proportion of simple calls decreases to over 30 percent of the bout. In this paper, we are interested in the proportion of calls prior to the moment when the females get close.

According to the studies of M. Ryan in Barro Colorado, Panama (\cite{ryan1985tungara}), the reproductive ritual takes place from 19:00 to 24:00. The ritual comprises the call bouts, the assessment and choice of the female, the mating (amplexus) and the nesting. The males wait from the shore, separated at least by $5$ $cm$ from one another, for the females in the pond. Males produce a sound called \textit{meu} (see \cite{ryan1985tungara}) to prevent other males from getting closer. If a male gets close enough to another male, he is pushed away. Once a female has come close enough to several choruses and makes a choice, the chosen male gets on the female back putting together their cloacae. The female transfer her eggs into the male cloaca for the male to fertilize them. In this process, the female transports the male on her back away from the shoreline of the pond. After midnight, the couple comes back to the the shoreline of the pond to build a foam nest that contains over 230 fertilized eggs \cite{ryan1985tungara}. This implies that once a male is chosen, it does not mate again. T\'ungara males do not fight between one another to mate, instead females walk towards them. In \cite{ryan1980female} it was observed that non-calling males sometimes intercept females that are going towards calling males. In consequence, this can be also considered a strategy in mating. In \cite{ryan1980female} it is also mentioned that it has not been observed unmated males displace a male in amplexus.

In \cite{ryan1985tungara}, Ryan argues that males do not come back to the same place in the pond every night. Furthermore, males do not defend any resource. This implies that the choice of the female is not oriented to obtain better resources. This is also evident due to the fact that females mostly build the nest in a place different from where they choose the males.

Females prefer complex calls than simple ones. The data in \cite{tarano2015choosing} accounts for this preference. In that study it is established that the probability of a female to choose a complex caller over a simple caller is around $p := 0.8$. Moreover, in \cite{tarano2015choosing} it is observed that the mean time for a female to cast a choice between a complex and a simple caller is not different from the mean time to cast a choice between two complex callers. In \cite{ryan1985tungara} it is also shown that the accuracy with which females detect either a complex or a simple call is the same. Therefore, females actually make a decision between simple and complex callers entirely based on the type of call. M. Ryan also showed that the bigger the chorus, the more females are attracted.

T\'ungara frogs have several predators. Among them, we find opossums \emph{Philandeu opossum} that detect the frogs because of the sound, the South American bullfrog \emph{Leptodactylus pentadactylus} that detects the frogs visually and the crab \emph{Potamocarcinus richmondi} that detects the frogs visually too. Regardless or these species, the main predator of T\'ungara frog is the Fringe-lipped bat \emph{Trachops cirrhosus} that detects the frog through echolocation. These bats identify the high frequency sounds more easily. M. Ryan and M. Tuttle proved that between a simple and a complex caller, bats choose the complex with probability $\pi:= \frac{2}{3}$ (see \cite{ryan2011replication}).  Frogs detect bats visually, and the effect of the presence of a bat in the choruses is that all frogs become quiet and the pause between bouts become longer (\cite{ryan1985tungara}). Regarding the response time of the bats, in \cite{page2008effect} it is shown that the response of the bats to simple calls is longer than the response to complex calls.

Based on his studies, M. Ryan claims that the variety of calls in these amphibians has evolved because: ``it allows a male to adjust call complexity to effect a compromise between maximizing mate attraction ability and minimizing predation risk" (\cite{ryan1985tungara}). In consequence, we study the proportion of simple calls to complex calls in a bout, to explain the idea of maximizing reproductive success and minimizing predation risk. In this paper we propose that the dynamic of a replicator equation capture this phenomenon.

In Section 2, we present models with two and three strategies. The first one accounts for the simple call and complex call regarded as strategies of a symmetric game. In the second model we added the quiet state as an strategy. In Section 3, we present the results of the stability analysis of the equilibrium points of our models. We also extrapolate intervals for the parameters that cannot be found in the literature so that the model reproduces the reported observations. The justification of the results presented are exposed in an Appendix in the section 5.

\section{The models}

\subsection{Model with two strategies}
\label{secmodel2}

The payoff for the two strategies, complex call and simple call, will be measured in terms of their reproductive success. We will denote by $C$ and $S$ respectively the strategies of complex call and simple call. Our symmetric game is set as follows: we have two males in a pond using strategies $X,Y$, where $X,Y\in\{ C, S \}$, a female assessing them, and potentially nearby there is a bat considering to attack.

In order to capture the previous ideas, we define the parameters shown in Table \ref{parameters2} where by an encounter $XY$ we mean the two frogs are using strategies $X$ and $Y$.

\begin{table}[H]
\centering
  \begin{adjustbox}{max width=\textwidth}

   \begin{tabular}{cccc}
    \toprule 
Parameter & Description\\  \midrule

\rowcolor[gray]{.9}  $p$ & Probability that a female chooses a complex caller in an encounter $CS$ \\
 
 $\pi$ & Probability that a bat chooses a complex caller in an encounter $CS$ \\
 
 \rowcolor[gray]{.9} $r_{XY}$ & Average number of choices that a female makes per unit of time in an encounter $XY$ \\
 
 $\rho_{XY}$ & Average number of choices that a bat makes in an encounter $XY$ \\
 
 \rowcolor[gray]{.9} $d_S $ & Ratio of $r_{SS}$ to $r_{CS}$ ($r_{SS}=d_Sr_{CS}$) \\
 
 $d_C$ & Ratio of $r_{CC}$ to $r_{CS}$ ($r_{CC}=d_Cr_{CS}$) \\
 
 \rowcolor[gray]{.9} $\delta_S$ & Ratio of $\rho_{SS}$ to $\rho_{CS}$ ($ \rho_{SS}=\delta_S\rho_{CS}$) \\ 
 $\delta_C $ & Ratio of $\rho_{CC}$ to $\rho_{CS}$ ($ \rho_{CC}=\delta_C\rho_{CS}$) \\

    \bottomrule
    \end{tabular}
\end{adjustbox}
\caption{Parameters of the model with two strategies}
\label{parameters2}

\end{table}

We denote by $P(X,Y)$ the payoff of a male using the strategy $X$ in an encounter $XY$. Therefore we have
$$
\begin{array}{rclcrcl}
P(S,S) & = & \dfrac{0.5r_{SS}}{0.5\rho_{SS}} & & P(S,C) & = & \dfrac{(1-p)r_{CS}}{(1-\pi)\rho_{CS}}\\
 & & & & & & \\
P(C,S) & = & \dfrac{p r_{CS}}{\pi \rho_{CS}} & & P(C,C) & = & \dfrac{0.5r_{CC}}{0.5\rho_{CC}}

\end{array}
$$
For example, to obtain the payoff $P(C,S)$ we proceed as follows. In an encounter $CS$, the average number of decisions cast by a female per unit of time is $r_{CS}$ and $p$ of them will be in favor of the complex caller. Similarly, in an encounter $CS$, the average number of decisions cast by a bat per unit of time is $\rho_{CS}$ and $\pi$ of them will be for the complex caller; thus, the expected time a frog can play the $C$ strategy in $CS$ encounters is $(\pi\rho_{CS})^{-1}$. The payoff is similarly defined for the rest of the encounters. The probability that a female frog or a bat chooses one or the other is $1/2$ if both male frogs are using the same strategy. Using the definition of $d_S,d_C,\rho_s,\rho_c$ given in Table \ref{parameters2},  we  get the payoff matrix in (\ref{game2}). \\

\begin{equation}\label{game2}
A_2 =  \begin{pmatrix}

 P(S,S) & P(S,C) \\
 P(C,S) & P(C,C)\\
 \end{pmatrix}= 
\frac{r_{CS}}{\rho_{CS}} \begin{pmatrix}

 \frac{d_S}{\delta_S} & \frac{(1-p)}{(1-\pi)} \\
 \frac{p}{\pi} &\frac{d_C}{\delta_C}\\

\end{pmatrix}
\end{equation}

\subsection{Model with three strategies} \label{secmodel3}

As we mentioned in the introduction, quiet males mate by intercepting females that were walking towards calling males. The quiet strategy leads males to be safe from predation. However, this strategy reduces the chances of males of being chosen by females. In an encounter between a quiet and a calling male, let $\theta$ be the probability that the quiet frog intercepts a female that goes towards the calling male.  Let $Q$, $S$ and $C$ denote respectively the strategies of staying quiet, producing simple calls and producing complex calls. In Table \ref{parameters3}, we define the parameters of the model.

\begin{table}[H]
\centering
  \begin{adjustbox}{max width=\textwidth}

   \begin{tabular}{cccc}
    \toprule 
Parameter & Description\\  \midrule

\rowcolor[gray]{.9} $\theta$ &  Probability that a quiet male intercepts a female going toward a calling male \\

 $p$ & Probability that a female chooses a complex caller in an encounter $CS$ \\

\rowcolor[gray]{.9} $\pi$ & Probability that a bat chooses a complex caller in an encounter $CS$ \\

 $r_{XY}$ & Average number of choices that a female makes per unit of time in an encounter $XY$  \\
 
\rowcolor[gray]{.9} $\rho_{XY}$ & Average number of choices that a bat makes per unit of time in an encounter $XY$  \\

 $d_{XY} $ & Ratio of $r_{XY}$ to $r_{CS}$ ($r_{XY} = d_{XY} r_{CS}$), for $XY \in \{SQ,CQ\}$ \\
 
 \rowcolor[gray]{.9}$d_{X} $ & Ratio of $r_{XX}$ to $r_{CS}$ ($r_{XX} = d_{X} r_{CS}$), for $X \in \{S,C\}$ \\
 
 $\delta_{XY} $ & Ratio of $\rho_{XY}$ to $\rho_{CS}$ ($ \rho_{XY} = \delta_{XY} \rho_{CS}$), for $XY \in \{SQ,CQ\}$ \\
 
\rowcolor[gray]{.9} $\delta_{X} $ & Ratio of $\rho_{XX}$ to $\rho_{CS}$ ($ \rho_{XX} = \delta_{X} \rho_{CS}$), for $X \in \{S,C\}$ \\

 $R$ & Ratio of $r_{XQ}$ to $r_{XS}$: reduction in number of choices a female makes per unit of time\\
  & when a quiet male takes the place of a simple caller ($r_{SQ} = R r_{SS}$ and $r_{CQ} = R r_{CS}$) \\

    \bottomrule
    \end{tabular}
\end{adjustbox}
\caption{Parameters of the model with three strategies}
\label{parameters3}
\end{table}

The payoffs $P(X,Y)$ for the strategy $X$ in each encounter $XY$ are the following:

$$
\begin{array}{rclcrclcrcl}
P(Q,Q) & = & 0 & &P(Q,S) & = & \theta \dfrac{ r_{SQ}}{ \rho_{SQ}} & & P(Q,C) & = &\theta \dfrac{r_{CQ}}{\rho_{CQ}}\\
 & & & & & & \\
P(S,Q) & = & (1-\theta) \dfrac{ r_{SQ}}{ \rho_{SQ}}& &P(S,S) & = & \dfrac{0.5r_{SS}}{0.5\rho_{SS}} & & P(S,C) & = & \dfrac{(1-p)r_{CS}}{(1-\pi)\rho_{CS}}\\
 & & & & & & \\
P(C,Q) & = & (1-\theta) \dfrac{r_{CQ}}{\rho_{CQ}} & &P(C,S) & = & \dfrac{p r_{CS}}{\pi \rho_{CS}} & & P(C,C) & = & \dfrac{0.5r_{CC}}{0.5\rho_{CC}}
\end{array}
$$

Payoffs are obtained as in the two strategies scenarios. For encounters involving a quiet frog we reason as follows. Whenever there is a $XQ$ encounter, for $X\in\{S,C\}$, the female and the bat will choose the calling frog with probability $1$; therefore, the expected number of times the calling frog will be chosen by a female is $r_{XQ}/\rho_{XQ}$. Now, in those encounters the quiet frog will only get a payoff when he manages to intercept the female frog. This happens with a ratio $\theta$.

In \cite{ryan1985tungara}, M. Ryan showed that females are more lured to bigger choruses. Therefore, we assume that, for $X\in\{S,C\}$, the rate at which a female chooses in a $XQ$ encounter is less than in a $XS$ encounter. Let $R$ be this reduction rate, so that $r_{SQ} = R r_{SS}$ and $r_{CQ} = r_{CS}$. From \cite{ryan1985tungara}, we estimate that $R=0.8$.

Note that $d_{SQ} = r_{SQ}/r_{SC} =(\frac{r_{SQ}}{r_{SS}})(\frac{r_{SS}}{r_{SC}})= R d_{S}$ and $d_{CQ} = \frac{r_{CQ}}{r_{CS}} = R$.

In \cite{halfwerk2014risky}, W. Halfwerk \emph{et al}. showed that bats mainly locate their preys due to high frequency sounds. Therefore, we assume that, for $X\in\{S,C\}$, the rate at which a bat cast his choice in a $XQ$ encounter is the same as in a $XS$ encounter. Hence, $\rho_{SQ}=\rho_{SS}$ and $\rho_{CQ}=\rho_{CS}$.

Note that $\delta_{SQ} = \rho_{SQ}/\rho_{CS}= (\frac{\rho_{SQ}}{\rho_{SS}})(\frac{\rho_{SS}}{\rho_{CS}})= \delta_{S}$ and $\delta_{CQ} = \frac{\rho_{CQ}}{\rho_{CS}} =1$.

In Table \ref{assump3}, we summarize these assumptions.
\begin{table}[H]
\centering
  \begin{adjustbox}{max width=\textwidth}

   \begin{tabular}{cccc}
    \toprule 

 Assumption & Reference\\ \midrule
 \hline
 
\rowcolor[gray]{.9} $P(Q,Q)=0$ & \cite{ryan1985tungara} \\
 
$r_{SQ} = R r_{SS}$ and $r_{CQ} = r_{CS}$  &  \cite{ryan1985tungara} \\

\rowcolor[gray]{.9} $\rho_{SQ}=\rho_{SS}$ and $\rho_{CQ}=\rho_{CS}$ & \cite{halfwerk2014risky}\\

    \bottomrule
    \end{tabular}
\end{adjustbox}
\caption{Assumptions on the quiet strategy}
\label{assump3}

\end{table}

Using the parameters in Table \ref{parameters3} and the assumptions in Table \ref{assump3},  we get the payoff matrix in \ref{game3}.  

\begin{equation}\label{game3}
A_3=\begin{pmatrix}

P(Q,Q) & P(Q,S) & P(Q,C) \\   
P(S,Q) & P(S,S) & P(S,C) \\
P(C,Q) & P(C,S) & P(C,C) \\

\end{pmatrix} =
\frac{r_{CS}}{\rho_{CS}} \begin{pmatrix}

0 & \theta R \frac{d_S}{\delta_S} & \theta R \\   
(1-\theta) R \frac{d_S}{\delta_S} & \frac{d_S}{\delta_S} & \frac{(1-p)}{(1-\pi)} \\
(1-\theta) R & \frac{p}{\pi} &\frac{d_C}{\delta_C}\\

\end{pmatrix}.
\end{equation}

\section{Results}

\subsection{Model with two strategies} \label{secresults2}

Let us remark that in a symmetric game with two strategies the stable points of the replicator equation are exactly the \textit{evolutionary stable strategies} (ESS) (see \cite[Proposition 3.15]{weibull1997evolutionary}). Therefore, the optimal strategies correspond to the stable points of the replicator equation defined by the matrix in (\ref{game2}). By optimal we mean that these strategies can not be invaded by others.

The replicator equation for the proportion of simple calls $s$ in our model with two strategies is (see Subsection \ref{apequilibria2} in Appendix):


\begin{equation}
\dot{s} = s(1-s)(a-(a+b)s) 
\label{replicator2s}
\end{equation}
where  $$a := ( \frac{1-p}{1- \pi}-\frac{d_C}{ \delta_C})\frac{r_{CS}}{\rho_{CS}}\quad\textrm{and}\quad b := (\frac{p}{ \pi} -\frac{d_S}{ \delta_S})\frac{r_{CS}}{\rho_{CS}}.$$ The equilibrium points are $s_0=0, s_1=1, \hat{s}=\frac{a}{a+b}$. 
 
From \cite{tarano2015choosing}, the rate for which a female makes a choice in a $CS$ encounter and in a $CC$ encounter is the same. Therefore, we take $d_C =1$.

Because of the preference of the bats for complex callers, we take $\delta_C>1>\delta_S$. Similarly, because of the preference of the females for complex callers, we will assume that $r_{SC}\ge r_{SS}$. In consequence, we take $d_S < 1$.

\begin{figure}[H]
\centering
\includegraphics[scale = 0.7]{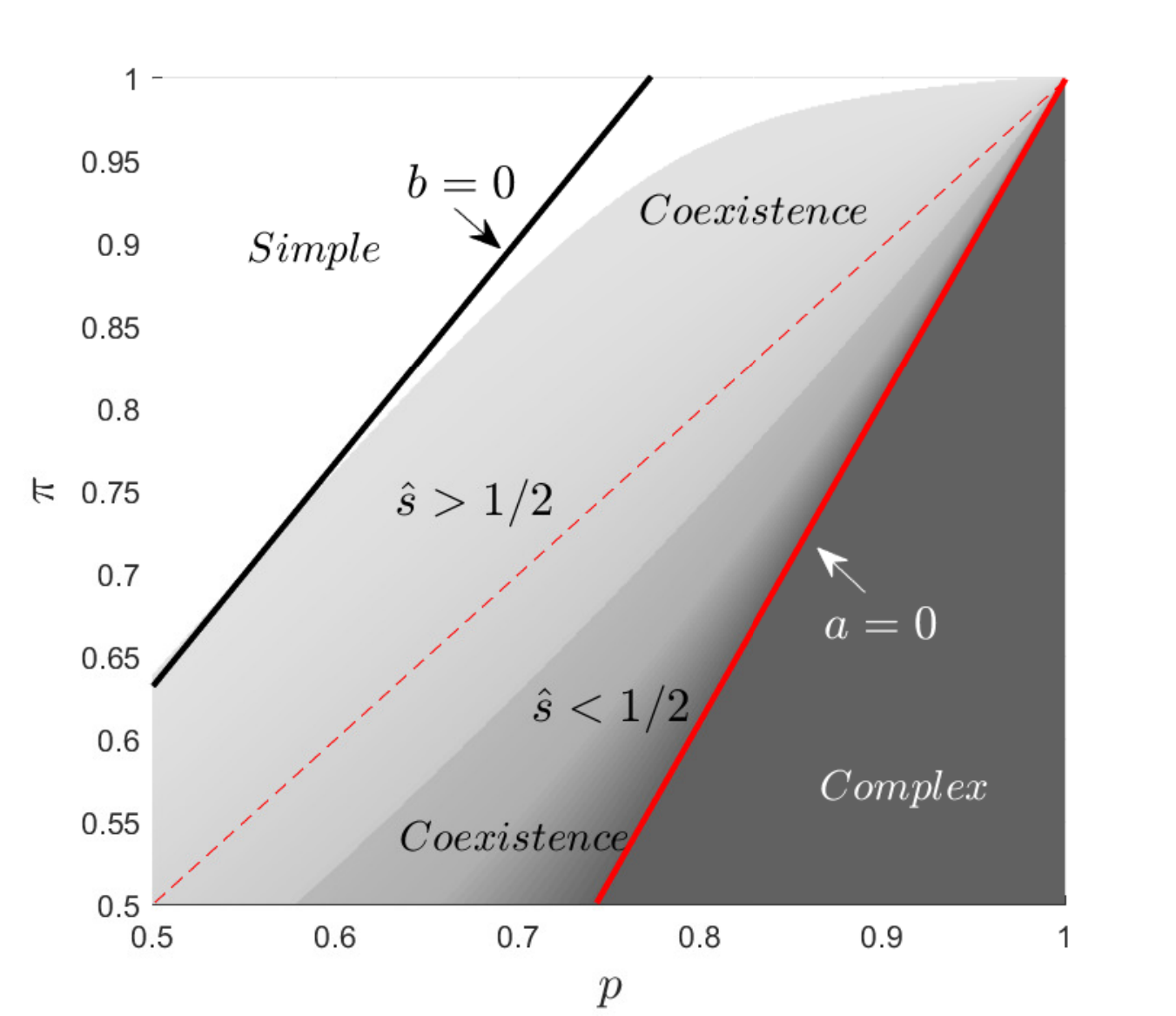}
\caption{Stability regions for the equilibria $\hat{s}=0, \hat{s}=1, \hat{s}=\frac{a}{a+b}$ in the $p\pi$-plane. The pure strategies $s_0=0$ and $s_1=1$ are respectively stable above the $b=0$ line and below the $a=0$ line; in-between there is coexistence. It is reported in \cite{ryan2011replication} and in \cite{tarano2015choosing} that $\pi > p$ (the region below the dashed red line). It is also reported in \cite{bernal2007cues} that there exists coexistence and $\hat{s} > 1/2$. Therefore, the parameters $p$ and $\pi$ should be between the curve $\hat{s}=1/2$ and the red line $a=0$}
\label{regions}

\end{figure}

Using the criteria shown in Table \ref{criteria} in Subsection \ref{apstability2} of Appendix, in the $p\pi$-plane, the lines $a=0$ or $b=0$  subdivide the unit square in regions of stability for each of the three equilibria (Figure \ref{regions}). The explicit equation for the lines $a=0$ and $b=0$ are respectively (\ref{eql1}) and (\ref{eql2}) in Subsection \ref{apstability2} of Appendix.

From the literature, we know that the preference of females for complex callers over simple callers is greater than the preference of bats for complex callers over simple callers. Therefore $p > \pi$. Moreover, both strategies coexist in a proportion $0<\hat{s}<1$ of simple calls. In consequence, the values $p$ and $\pi$ are in the region between $a=0$ and $b=0$.

We also have that the lines $a=0$ and $b=0$ do not intersect in the unit square and $a=0$ is below $b=0$. Otherwise, we would have an scenario that makes no biological sense, as it is explained in Subsection \ref{apstability2} of Appendix. Similarly, from a biological perspective, both slopes must be greater than 1, which means that in the parameter of the model we have $\delta_S\geq d_S$ and $\delta_C\geq d_C$.

\begin{figure}[H]
 \centering
\includegraphics[scale = 0.6]{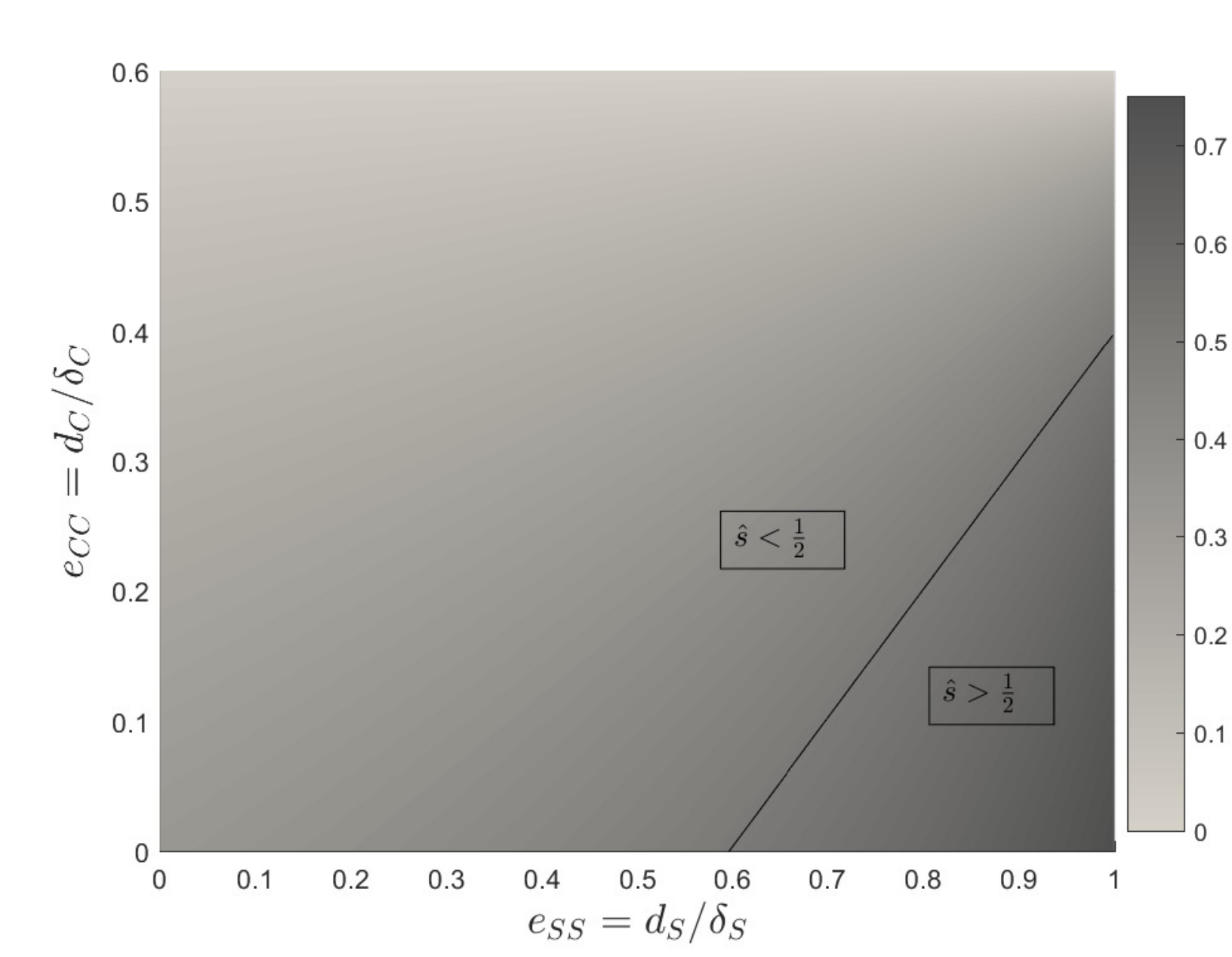}
\caption{Using the values of $p$ and $\pi$ found in the literature we get the values of $\hat{s}$ shown in this figure. Is is reported in literature the the value of $\hat{s}$ is greater than $1/2$, which is the case for the values of $e_{SS}$ and $e_{CC}$ below the black line in this figure}
\label{fxssxcc2}
\end{figure}

We define $e_{SC} = (1-p)/(1-\pi)$ and $e_{SC} = p/\pi$, and
we assume that $p=4/5$ and $\pi = 2/3$. Figure \ref{fxssxcc2} also shows the line $\hat{s}=1/2$ (see  the equation (\ref{line0.5}) in Subsection \ref{apstability2} of Appendix). Below this line we have the values of $e_{SS}$ and $e_{CC}$ for which $\hat{s} >1/2$, as it is reported in literature (see \cite{bernal2007cues}).  

\subsection{Model with three strategies} 
\label{secresults3}

If $\theta= 0$ males do not intercept females, and we recover the same dynamics as in the model with two strategies. Using the matrix $A_3$ introduced in \ref{game3}, we have that the replicator equation is given by the following system of equations:

\begin{equation}
\begin{cases}
\dot{q} =q((A_3 \bar{x})_1 - \bar{x}^t A_3 \bar{x})\\
\dot{s} =s((A_3 \bar{x})_2 - \bar{x}^t A_3 \bar{x})\\

\end{cases}
\label{replicator3}
\end{equation}

where $q$ and $s$ are respectively the proportion of quiet males and complex callers, $\bar{x}$ is the vector $[q, s, 1-q-s]$ and $(A_3 \bar{x})_i$ is the entry $i$ of the vector $A_3 \bar{x}$ for $i=1,2$.

In Subsection \ref{apequilibria3} of Appendix we show that there are at most three equilibria besides the pure strategies $(q,s)=(1,0), (0,1)$ and $(0,0)$. Two of them are $w_{QC}$, with $s=0$, and $w_{SC}$, with $q =0$. A third possible equilibrium is $w^*=(q^*,s^*)$ where the three strategies coexist, $q^*,s^*>0$ and $q^*+s^*<1$.

We define $e_{XY} = P(X,Y) (\frac{\rho_{XY}}{r_{XY}})$ for any encounter $XY$ (see the matrices in \ref{game3}). For example, $e_{SC} = (1-p)/(1-\pi)$ and $e_{SC} = p/\pi$.

\begin{figure}[H]
\centering
 \begin{tabular}{cc}

   \includegraphics[scale=0.5]{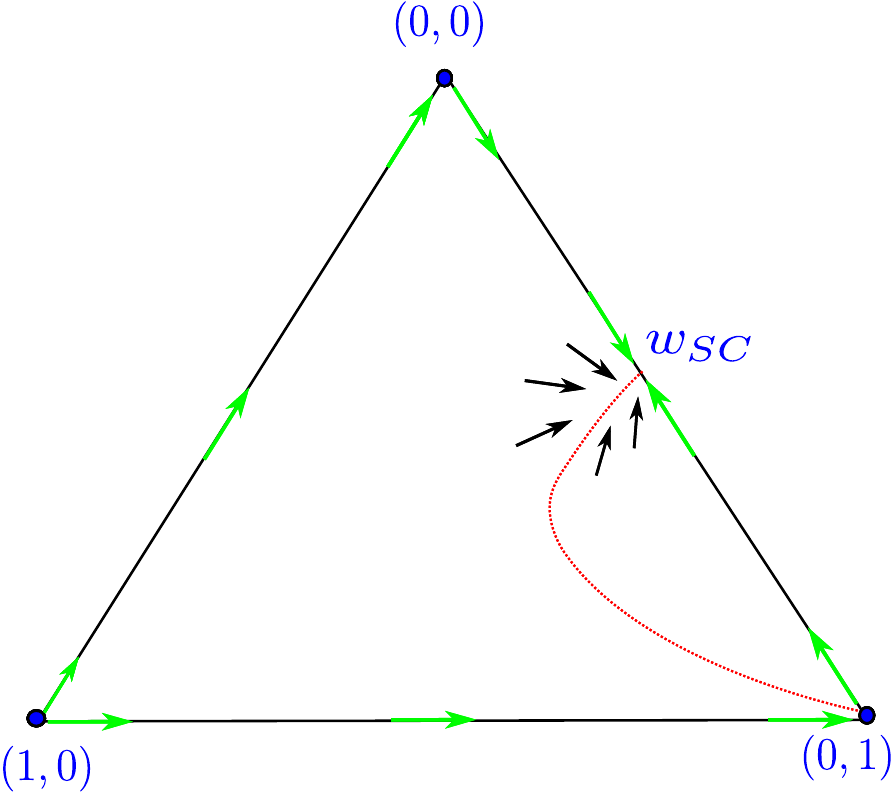} &
    \includegraphics[scale=0.5]{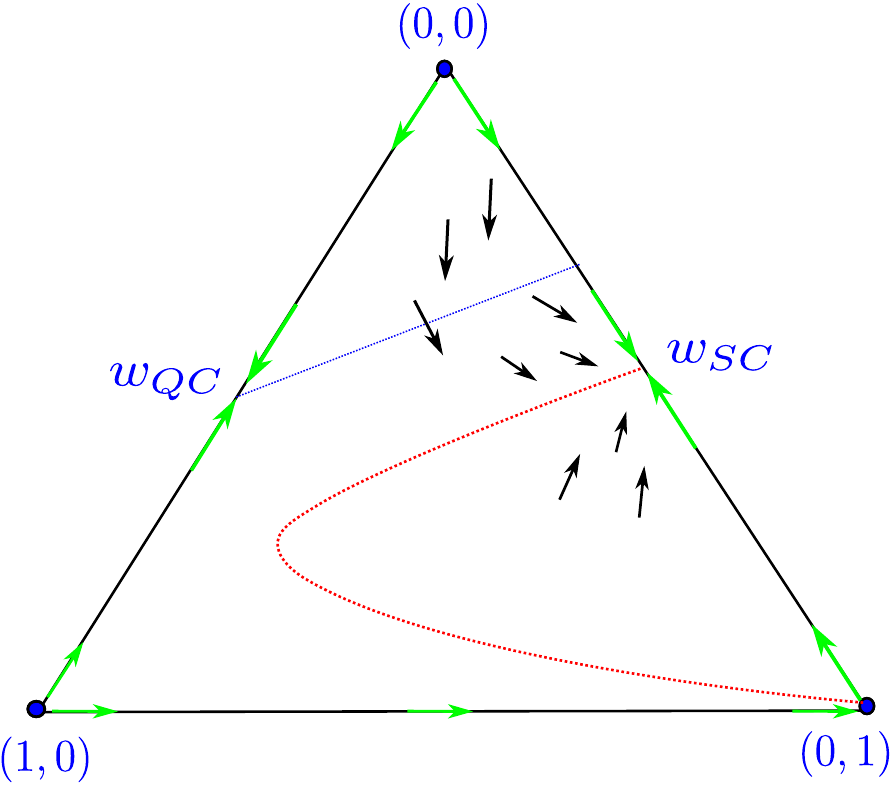}  \\
     
    $(a)$ $0<\theta < \frac{e_{CC}}{R}$ & $(b)$ $\frac{e_{CC}}{R}<\theta < \frac{e_{SC}}{R}$\\
 
 \end{tabular}

  \caption{The red dashed line represents the nullcline $\dot{s} =0$ and the blue dashed line represents the nullcline $\dot{q} =0$. The black arrows represent the directions of the trajectories of the model. In each figure we assume that $w_{SC}$ exists, which is equivalent to $e_{CC} <e_{SC}$. Figure $(a)$ shows that if $0<\theta < \frac{e_{CC}}{R}$, we obtain the behavior represented by the two strategies model. Figure $(b)$ shows that if  $\frac{e_{CC}}{R}<\theta < \frac{e_{SC}}{R}$, the equilibrium $w_{QC}$ appears. In the scenarios $(a)$ and $(b)$, $w_{SC}$ is the only stable equilibria.} 
   \label{corner3}
\end{figure}

From the eigenvalues and eigenvectors analysis in Subsection \ref{apstability3} of Appendix, we obtain that the equilibrium $w_{SC}$ exists if $e_{CC} < e_{SC}=0.6$. We assume that $e_{CC}$ always satisfies this inequality, as in the coexistence scenario of the model with two strategies. If $0<\theta < e_{CC}/R$, we have that $w_{QS}$ and $w^*$ do not exist. In this case, $w_{SC}$ is stable. In the subspace where $q+s=1$ we always have that the equilibrium $(0,1)$ is stable. In the subspace where $q=0$ we always have that $w_{SC}$ is stable. In the subspace where $s=0$, for $0<\theta < e_{CC}/R$ we have that $(0,0)$ is stable. If $e_{CC}/R<\theta < e_{SC}/R$, we have that $w^*$ does not exist and $w_{QS}$ exists. In this case, $w_{SC}$ is also stable, and the equilibrium $w_{QS}$ is stable in the subspace where $s=0$. Figure \ref{corner3} shows the previous two scenarios.

\begin{figure}[H]
\centering
 \begin{tabular}{ccc}

   \includegraphics[scale=0.45]{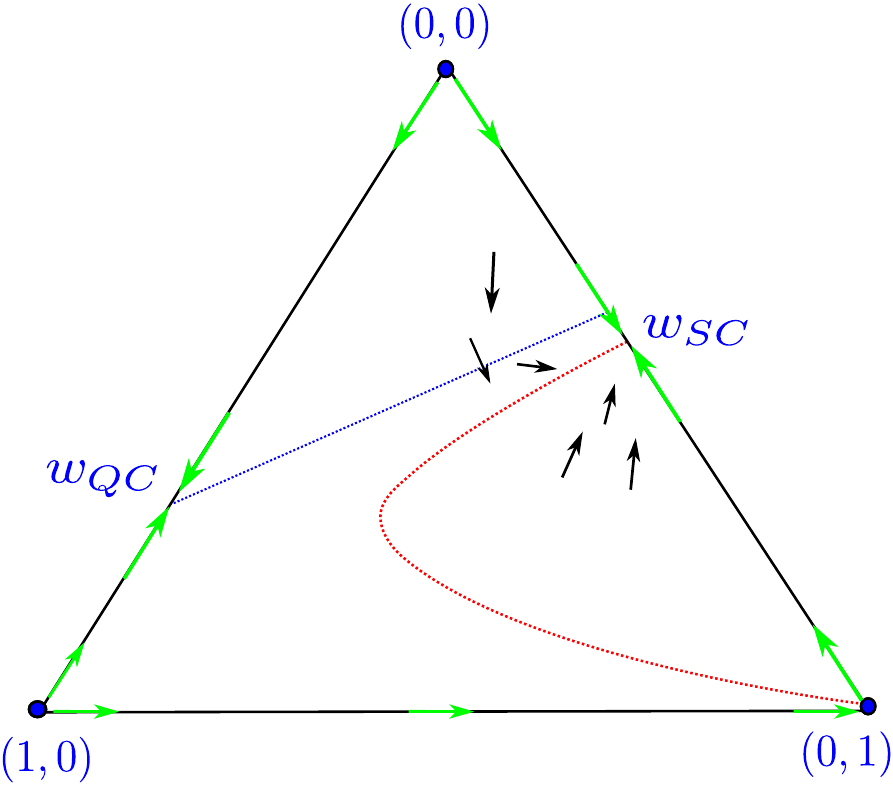} &
    \includegraphics[scale=0.45]{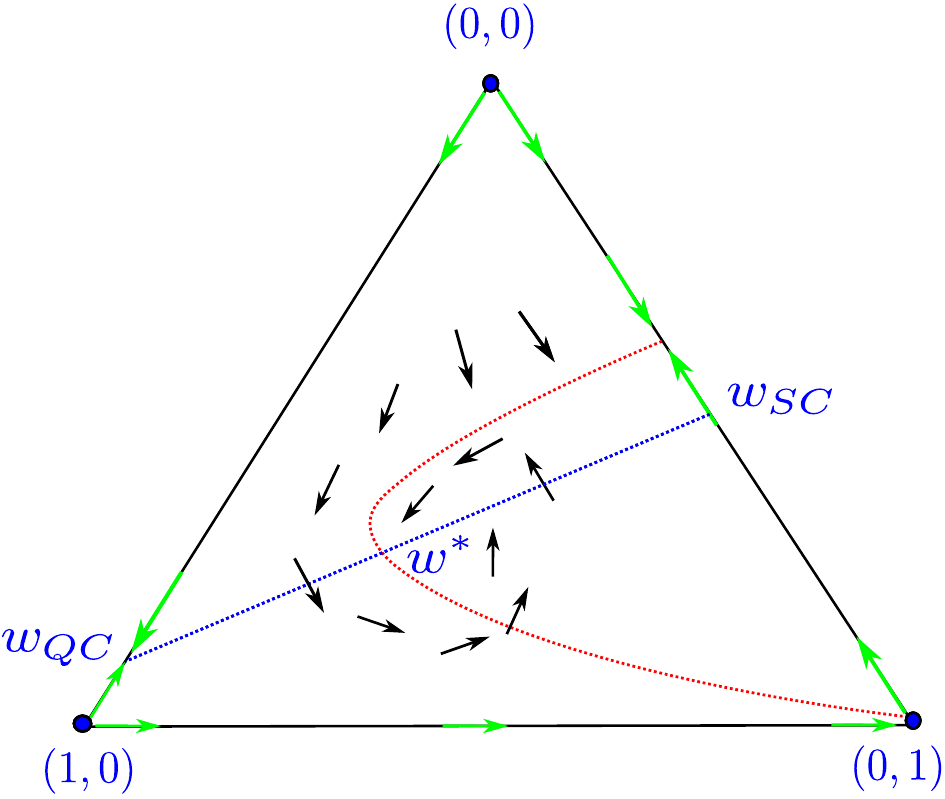} &
    \includegraphics[scale=0.45]{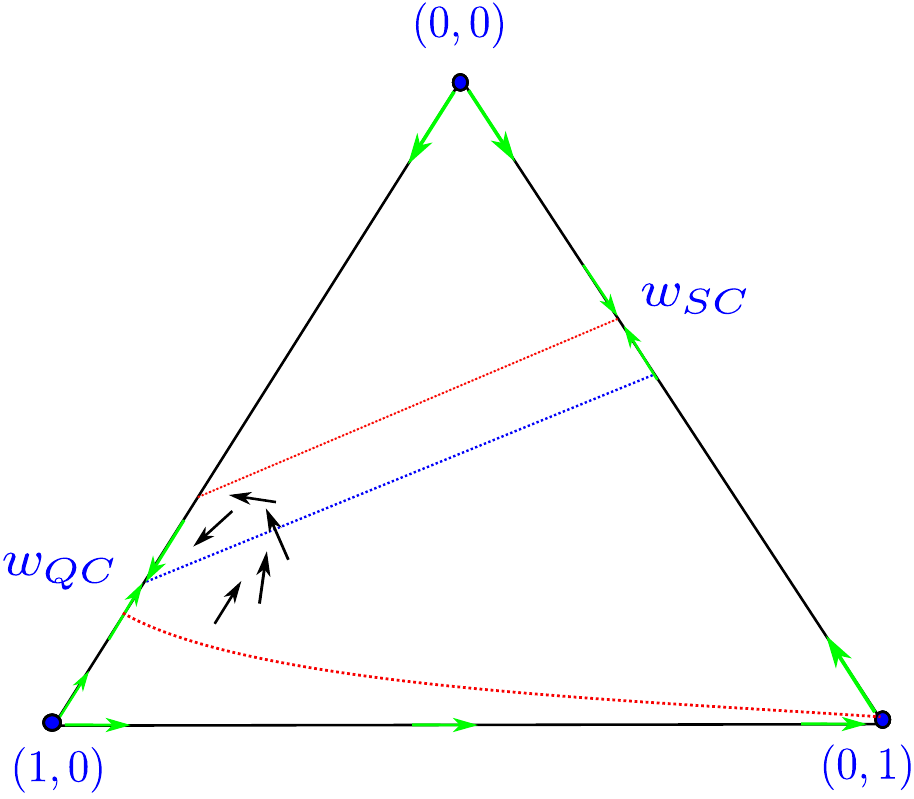} \\
      \includegraphics[scale=0.2]{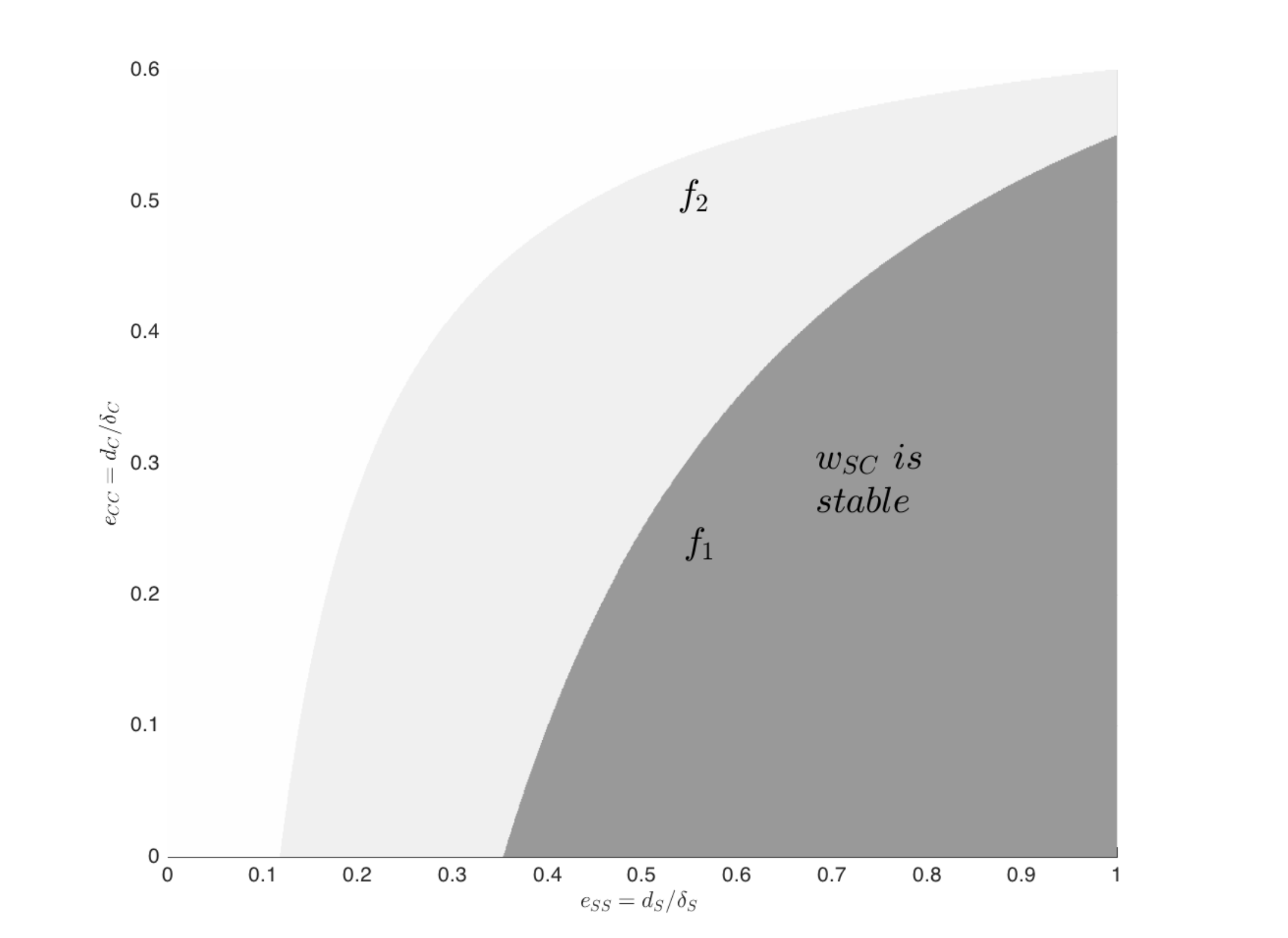} &
    \includegraphics[scale=0.2]{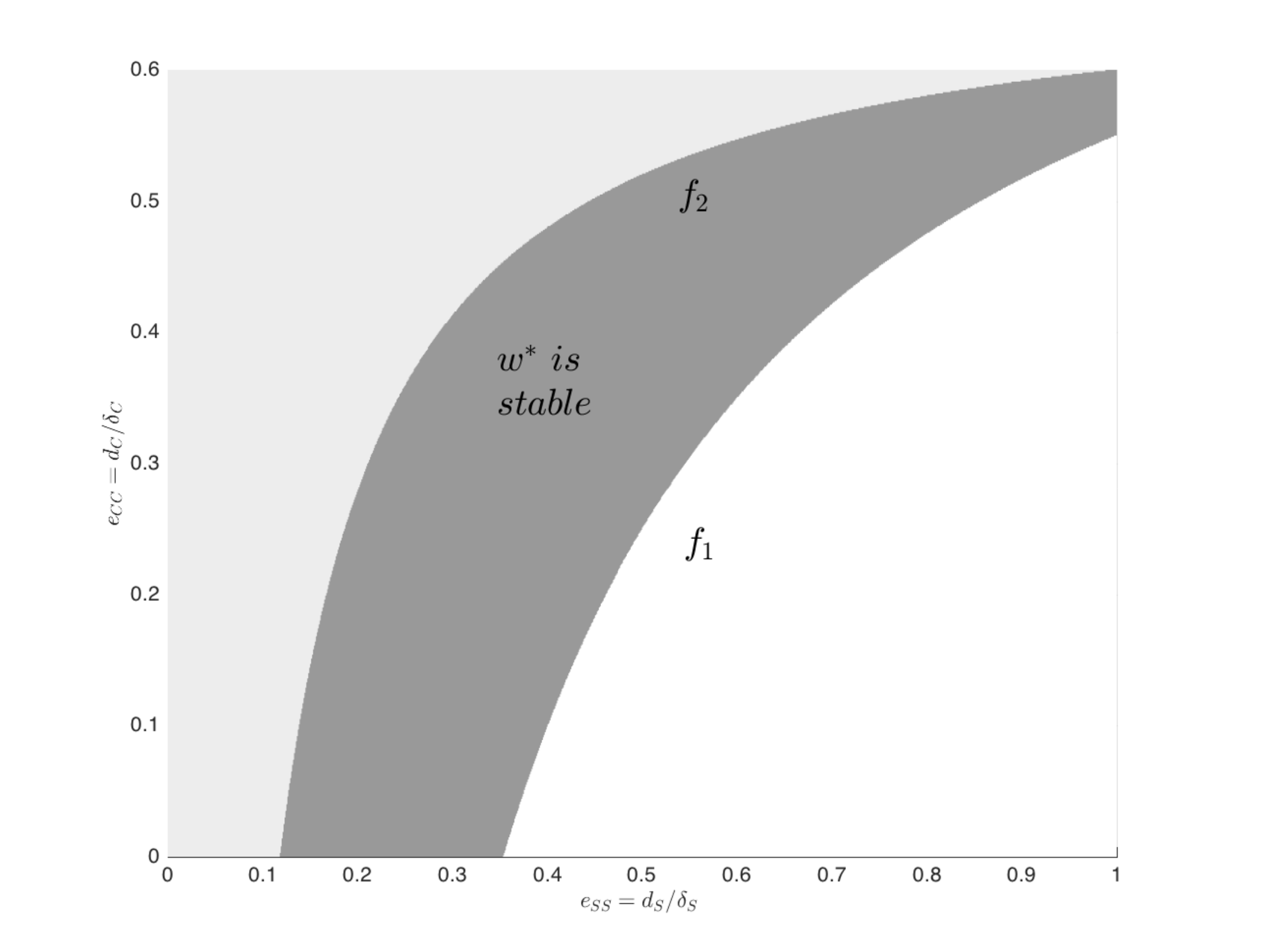} &
    \includegraphics[scale=0.2]{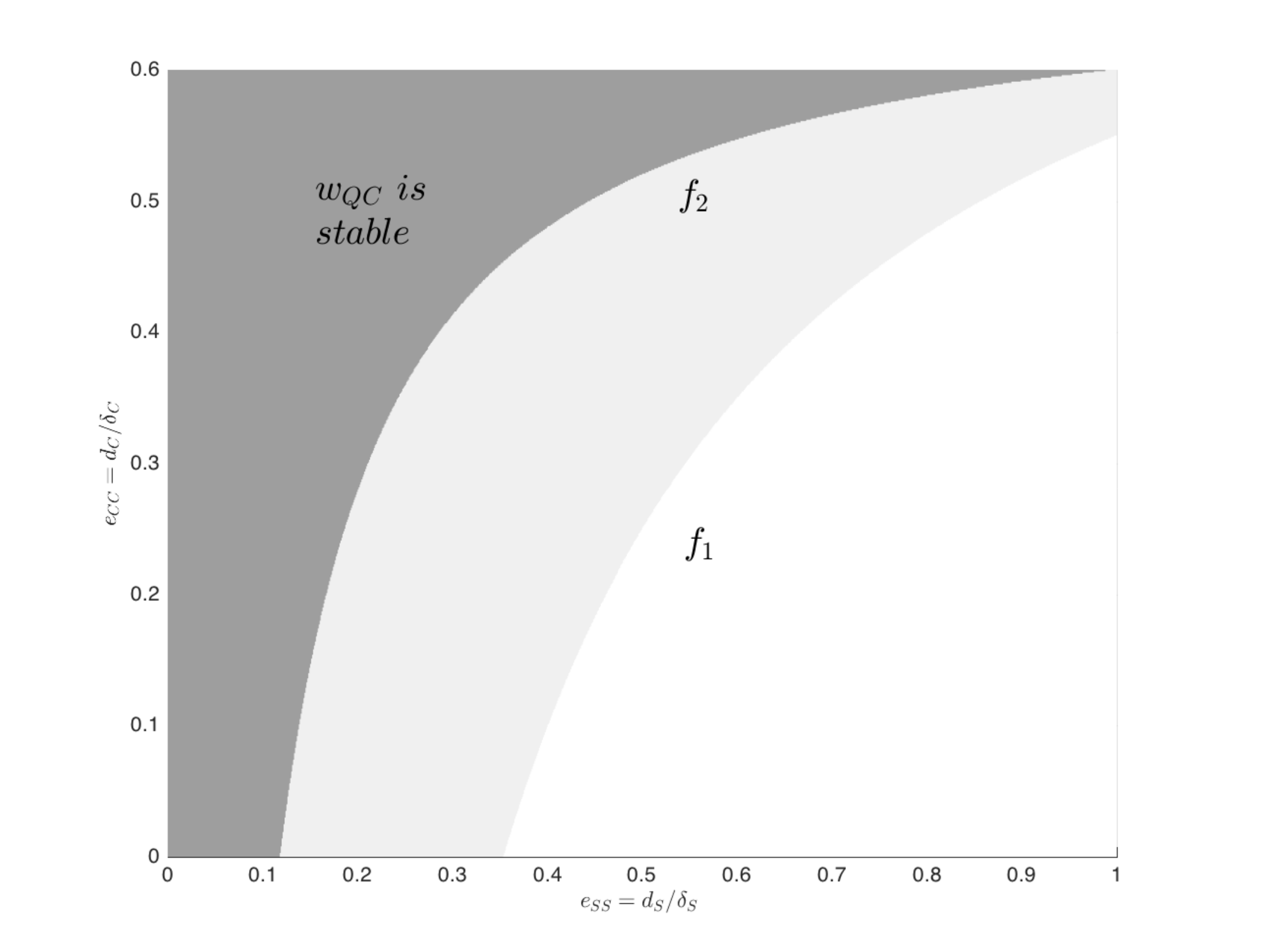} \\

 $(a)$ $ e_{CC} < f_1$  &$(b)$ $f_1 \leq e_{CC} \leq f_2$&$(c)$  $e_{CC} >f_2$\\
 \end{tabular}

  \caption{The red dashed line represents the nullcline $\dot{s} =0$ and the blue dashed line represents the nullcline $\dot{q} =0$. The black arrows represent the directions of the trajectories of the model. In each figure we assume that $w_{SC}$ exists, which is equivalent to $e_{CC} <e_{SC}=0.6$. We also assume that $ \frac{e_{SC}}{R}<\theta<1$. Figure $(a)$ shows that in the region $e_{CC}<f_1$ in the plane $e_{SS}e_{CC}$ (at the bottom), the equilibrium $w_{SC}$ is stable. Figure $(b)$ shows that in the region $f_1<e_{CC}<f_2$ in the plane $e_{SS}e_{CC}$ (at the bottom), the coexistence equilibrium $w^*$ appears and is stable. Similarly, Figure $(c)$ shows that in the region $e_{CC}>f_2$ in the plane $e_{SS}e_{CC}$ (at the bottom), the equilibrium $w_{QC}$ is stable. Furthermore, if we use the values of $e_{SS}$ and $e_{CC}$ for which $\hat{s}>1/2$ in the model of two strategies (see Figure \ref{fxssxcc2}), we have that $w_{QC}$ is never stable.} 

 \label{corner4}
\end{figure}

If $3/4 = e_{SC}/R<\theta<1$, we have that the coexistence equilibrium $w^*$ could exist (see the equation (\ref{coex3eq4}) in Subsection \ref{apequilibria3} of Appendix). We analyzed the existence of this equilibrium in terms of the parameters $e_{SS}$, $e_{CC}$ and $\theta$. Figure \ref{corner4} shows the regions in the plane $e_{SS}e_{CC}$ where there could be coexistence in terms of the functions $e_{CC}=f_1(\theta,e_{SS})$ and $e_{CC}=f_2(\theta, e_{SS})$ for fixed values of $\theta$ (see the equation (\ref{f1}) and the equation (\ref{f2}) in Subsection \ref{apstability3} of  Appendix for the definition of $f_1$ and $f_2$). For $\theta > e_{SC}/R$ we have that $w^*$ exists and is stable if $f_1 \leq e_{CC} \leq f_2$. However, $w^*$ does not exists if $ e_{CC} < f_1$ or $e_{CC} >f_2$. In these cases, $w_{SC}$ or $w_{QC}$ are respectively stable (see Figure \ref{corner4}). We have that the region of the plane $e_{SS}e_{CC}$ for which $\hat{s} > 1/2$ in the two strategies game (below the line of the Figure \ref{fxssxcc2} in Subsection \ref{secresults2}) is contained in the region below $f_2$ in Figure \ref{corner4} (see Subsection \ref{apstability3} of Appendix). Therefore, $w_{QC}$ would not be stable if we are based on the parameters of the model with two strategies.

\section{Discussion and conclusions}

Even though in \cite{lea2015irrationality} A. Lea and M. Ryan show that the decision of the female could be irrational, in this work we show that there is an scenario in which the behavior of the male T\'ungara frog is rational.
Regarding the model with two strategies, in Subsection \ref{secresults2} we concluded that
$\frac{\delta_C}{d_C} \geq 1$ and   $\frac{\delta_S}{d_S}\geq 1$ in order that the $ESS$ made sense when varying $p$ and $\pi$. Although we can suggest from \cite{tarano2015choosing} that $d_C$ on average is around $1$, we did not obtain from the literature the values of $\delta_C$, $\delta_S$ and $d_S$. Nonetheless, in \cite{page2008effect} there is data on the latency of the bat response to the complex call compared to the response to the simple call. In that study, it is shown that the response of the bat to the complex call is faster than the response of the bat to the simple call, which leads us to suggest that the rate of predation is larger as there are more complex callers (in particular, $\rho_{CS} > \rho_{CC}$). As $\delta_C=\frac{\rho_{CS}}{\rho_{CC}}$, we get that $\frac{\delta_C}{d_C} \geq 1$ is coherent. The latency times reported in \cite{page2008effect} are obtained setting a bat next to records of calls. However, our $\delta_S$ and $\delta_C$ measure the response of a bat that it is not close to an encounter (if a bat is close to the encounter, the males get quiet according to \cite{ryan1985tungara}). For this reason, although this data is evidence of difference in latency of the bat response, the values reported in \cite{page2008effect} are not considered as parameters of our models.
    
On the other hand, the preference of the female for the complex calls is over $0.8$ (\cite{ryan1985tungara} and \cite{tarano2015choosing}). This leads us to suggest that when there are no complex callers the response of the females decreases (in particular $d_S<1$). However, we did not find a study where the latency of the response of the female to only simple calls is measured. From the observations in \cite{page2008effect}, we also have that $\delta_S < 1$. Besides, the preference of the bats for the complex call is not as high as the preference of the female. Moreover, the sensitivity of the echolocation of the bats for the ripples made by the calls in the water (\cite{akre2011signal}) make these predator to be capable of responding to simple calls also. Therefore, it makes sense to assume that the reduction of the mating choice  of the female  is higher than the reduction of the predation rate of the bats when there are only simple call (which means $\frac{\delta_S}{d_S}\geq 1$).

The proportion of calls in a bout reported in \cite{bernal2007cues} is over 0.6 in a chorus in absence of bats and females. This natural behavior can be replicated by our two strategies model as it is shown in the Figure \ref{fxssxcc2} in Subsection \ref{secresults2} (below the line). From Figure \ref{fxssxcc2} we must have that $e_{SS} = \frac{d_S}{\delta_S}$ must be close to $1$  and $e_{CC} = \frac{d_C}{\delta_C}$ must be small. The fact that $\delta_C$ is large is suggesting that the predation rate is sensible to the complex callers that are added. This agrees with the fact that high frequency calls produce more conspicuous ripples in the water, which are easier to detect as it is shown in \cite{akre2011signal}. 

In consequence, the model is showing that there is a coherent scenario where the predation and reproductive conditions make possible to look at the proportion of calls reported in \cite{bernal2007cues} as an evolutionary stable strategy.

In \cite{ryan1980female}, it is observed that quiet males intercepts females that were walking towards a calling male. Hence, we could think this as an strategy to avoid predation and having reproductive success. Nonetheless, in \cite{ryan1985tungara} and \cite{bernal2007cues} it is observed that most of the males call in a bout. The aim of our second model was to look at the effect of adding  the quiet strategy to the first model. We obtained that the proportion $\theta$ of times that a quiet frog must intercept females to mate must be large in order to have that the quiet strategy is a rational option (in the sense that there is a stable point of the replicator equation in which the three strategies coexist). We obtained that the coexistence of the three strategies holds as long as $\theta > 3/4$. This means that a quiet frog should intercept females most of the times to get a reproductive success which justifies to keep quiet. Such a frequent behavior is not reported in the literature. In  \cite{ryan1980female} it is only mentioned that the quiet males sometimes intercept females that walk toward calling males, and in \cite{bernal2007cues} it is observed that in a bout the quiet strategy does not coexists with the other strategies. Therefore, our second model explains that the strategy observed in  \cite{ryan1980female} is not rational in a chorus in terms of reproductive success,  which agrees with the observed behavior in nature.

\section{Appendix}

\subsection{Model with two strategies: Equilibria of the model} \label{apequilibria2}
In Section \ref{secmodel2} we defined the following payoff matrix for the  game with two strategies:

$$A_2 =  
\frac{r_{CS}}{\rho_{CS}} \begin{pmatrix}

 \frac{d_S}{\delta_S} & \frac{(1-p)}{(1-\pi)} \\
 \frac{p}{\pi} &\frac{d_C}{\delta_C}\\

\end{pmatrix} =\frac{r_{CS}}{\rho_{CS}} \begin{pmatrix}
  
e_{SS} & e_{SC} \\
e_{CS} & e_{CC} \\

\end{pmatrix}. $$

From this matrix, we obtain the replicator equation:

$$
\dot{s} = s((A_2 \bar{x})_1 - \bar{x}^t A_2 \bar{x}). 
$$

This equation is equivalent to the equation: 

$$
\dot{s} = s(1-s)(a-(a+b)s). 
$$

where $$a := ( \frac{1-p}{1- \pi}-\frac{d_C}{ \delta_C})\frac{r_{CS}}{\rho_{CS}}\quad\textrm{and}\quad b := (\frac{p}{ \pi} -\frac{d_S}{ \delta_S})\frac{r_{CS}}{\rho_{CS}}.$$

We obtain three possible equilibria of this equation: the pure strategies $s_0=0$, $s_1=1$ and the coexistence equilibrium 
\begin{equation}
\hat{s} = \frac{a}{a+b}=\frac{e_{SC}-e_{CC}}{e_{SC}-e_{CC}+e_{CS}-e_{SS}}
\label{equilcoex2}
\end{equation}

if $0<\frac{a}{a+b}<1$.

\subsection{Model with two strategies: Stability of the equilibria} \label{apstability2}

Table \ref{criteria} shows the criteria for the stability of the equilibria presented in the previous subsection.

\begin{table}[H]
  \begin{center}
  
    \begin{tabular}{cc}
    \toprule 
    Equilibrium & Stability criteria \\ \midrule
        \rowcolor[gray]{.9}  $s_0=0$ & $a<0$  \\
  $s_1=1$  & $b<0$   \\

        \rowcolor[gray]{.9}  $\hat{s}=\frac{a}{a+b}$ & $\frac{-ab}{a+b}<0$ \\

    \bottomrule
    \end{tabular}
   \caption{Criteria for the stability of the equation (\ref{replicator2s}).}
     \label{criteria}
  \end{center}
\end{table}

We have that $ \frac{a}{a+b} = 0 $ if $0= \frac{1-p}{1- \pi}-\frac{d_C}{ \delta_C}$ or equivalently 

\begin{equation}
\pi = \frac{\delta_C}{d_C}p + (1-\frac{\delta_C}{d_C}).
\label{eql1}
\end{equation}

The equation  (\ref{eql1}) determines a straight line in the plane $p \pi$. Just below this line, we have that the simple strategy dominates. On the other hand, $\frac{a}{a+b} = 1$ if $0=\frac{p}{ \pi} -\frac{d_S}{ \delta_S}$ or equivalently 
\begin{equation}
\pi = \frac{\delta_S}{d_S}p.
\label{eql2}
\end{equation}

Just above the line defined by the equation (\ref{eql2}), the complex strategy dominates. Furthermore, both strategies coexist between the lines defined by the equations (\ref{eql1}) and (\ref{eql2}),  as Figure \ref{regions} shows.

In Subsection \ref{secresults2}, we had mentioned that the lines defined by the equations (\ref{eql1}) and (\ref{eql2}) must not cross each other for $ p = p^* < 1$. Otherwise, in the region between the lines for $p^*\leq p \leq 1$ we get that for fixed $p$, if $\pi$ increases, then $\hat{s}$ decreases. This would mean that if we fix the preference of the females, and we increase the preference of the bats for the complex calls, then an optimal strategy is to produce more complex calls, which does not make sense.

Therefore, the slope of the line $a=0$ can not be smaller than 1, which means that we must have $\frac{\delta_C}{d_C}\geq 1$ (see the equation (\ref{eql1})). Likewise, the slope of the line $b=0$ con not be below 1, which is equivalent to $\frac{\delta_S}{d_S}\geq 1$ (see the equation (\ref{eql2})).

From \cite{bernal2007cues} we have $\hat{s} \geq 0.5$. Therefore, using the equation (\ref{equilcoex2}) we have that $\hat{s} \geq 0.5$ if 
\begin{equation}
e_{CC} \leq e_{SS} - (e_{CS}-e_{SC})=e_{SS} -3/5.
\label{line0.5}
\end{equation}
This is the equation of the line in Figure \ref{fxssxcc2}.

\subsection{Model with three strategies: Equilibria of the model} \label{apequilibria3}

In Subsection \ref{secmodel3} we defined the following payoff matrix for the  game with three strategies:

$$ A_3=
\frac{r_{CS}}{\rho_{CS}} \begin{pmatrix}

0 & \theta R \frac{d_S}{\delta_S} & \theta R \\   
(1-\theta) R \frac{d_S}{\delta_S} & \frac{d_S}{\delta_S} & \frac{(1-p)}{(1-\pi)} \\
(1-\theta) R & \frac{p}{\pi} &\frac{d_C}{\delta_C}\\
\end{pmatrix}= \frac{r_{CS}}{\rho_{CS}} \begin{pmatrix}

0& e_{QS} &  e_{QC} \\   
e_{SQ} &e_{SS} & e_{SC} \\
e_{CQ} & e_{CS} & e_{CC} \\

\end{pmatrix}.  $$

We also defined the replicator equation given by:
$$ \begin{cases}
\dot{q} =q((A_3 \bar{x})_1 - \bar{x}^t A_3 \bar{x})\\
\dot{s} =s((A_3 \bar{x})_2 - \bar{x}^t A_3 \bar{x})\\

\end{cases} $$

We have that there are three equilibria in the corners of the triangle $\{(q,s): q+s\leq 1, q\geq 0, s\geq 0\}$, which are $(1,0)$, $(0,1)$ and $(0,0)$. The replicator equation has also the following equilibria:

\begin{equation}
w_{QC}:=(\frac{e_{QC}-e_{CC}}{e_{CQ}+e_{QC} - e_{CC} },0)
\label{equils0}
\end{equation}

\begin{equation}
w_{SC}:=(0, \frac{e_{SC}-e_{CC}}{e_{SC}-e_{CC} + e_{CS} - e_{SS}})
\label{equilq0}
\end{equation}

\begin{equation}
w_{QS}=( \frac{e_{QS}-e_{SS}}{e_{QS}-e_{SS} + e_{SQ}},  \frac{e_{QS}}{e_{QS}-e_{SS} + e_{SQ}})
\label{equilc0}
\end{equation}

We have that the equilibrium $w_{QC}$ is within the side of the triangle if $e_{QC} =R \theta > e_{CC}$. The equilibrium $w_{SC}$ is within the side of the triangle if $e_{SC} > e_{CC}$. However, the equilibrium $w_{QS}$ never exists within the side of the triangle because $e_{QS}-e_{SS}<0$ and $e_{SQ}>0$.

In addition, there could be a coexistence equilibrium of the three strategies $w^*$, which is defined as follows:

\begin{equation}
 w^*=(\frac{\nu}{\nu + \xi + \epsilon},\frac{\xi}{\nu+ \xi + \epsilon}) 
   \label{coex3eq4}
\end{equation}

where

\begin{equation}  
 \nu =  e_{CC}e_{SS} + e_{QS}e_{SC} + e_{QC}e_{CS} - e_{QS}e_{CC}- e_{QC}e_{SS} - e_{SC}e_{CS}
\label{coex3eq3}
\end{equation}

\begin{equation}
  \xi = e_{SQ}e_{QC}+e_{CQ}e_{SC}-e_{CQ}e_{QC}-e_{SQ}e_{CC}
  \label{coex3eq1}
  \end{equation}

\begin{equation}
  \epsilon =e_{CQ}e_{QC} + e_{SQ}e_{CC}  -e_{SQ}e_{QC}-e_{QC}e_{SC} 
    \label{coex3eq2}
    \end{equation}
  
\subsection{Model with three strategies: Stability of the equilibria} \label{apstability3} 

\begin{figure}[H]
\centering
\includegraphics[scale=0.5]{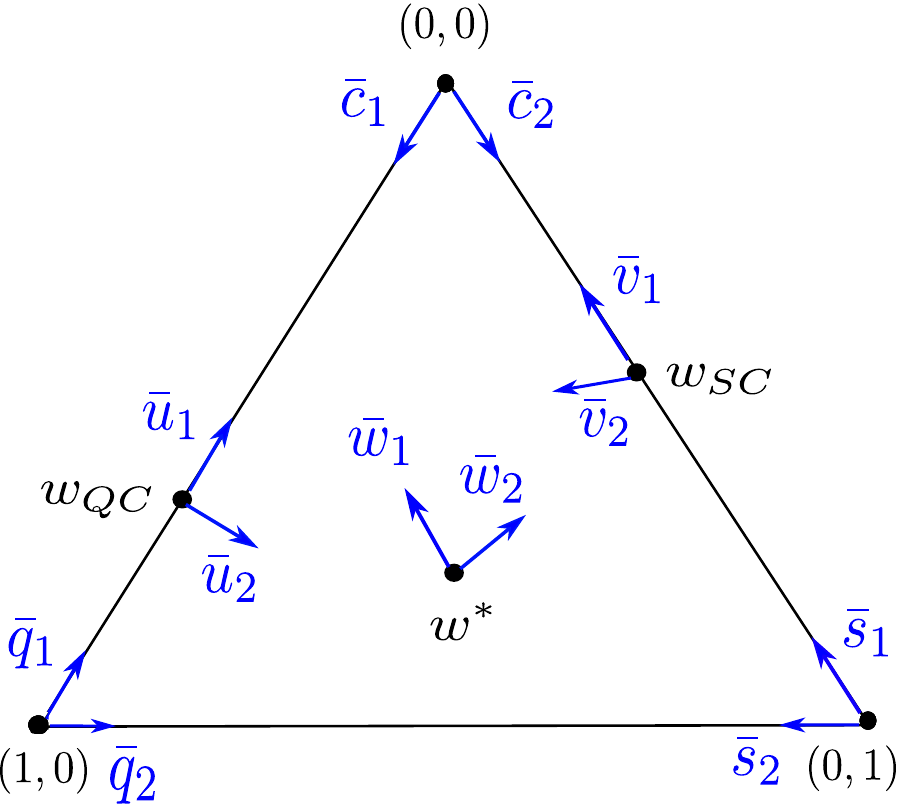}
\caption{Eigenvectors of equilibria.}
\label{vectors}
\end{figure}

Figure \ref{vectors} represents the eigenvectors for each equilibrium point and Table \ref{corners} shows their eigenvalues (except for $w^*$).

\begin{table}[H]
\centering
  \begin{adjustbox}{max width=\textwidth}

   \begin{tabular}{cccc}
    \toprule 
 Point & Eigenvector & Eigenvalue & Sign  \\ \midrule

\rowcolor[gray]{.9}$q=0, s=1$  & $\bar{s_1}=(0,-1)$ &
 $ e_{CS} - e_{SS}$ & + \\

 \rowcolor[gray]{.9}  & $\bar{s_2}=(1,-1)$ &
 $e_{QS}-e_{SS} = \theta R e_{SS} - e_{SS}$ & - \\

   $q=1, s=0$  & $\bar{q_1}=(1,-1)$ &
 $e_{SQ}$ & +\\

  & $\bar{q_2}=(-1,0)$ &
 $e_{CQ}$ & +\\

 \rowcolor[gray]{.9} $q=0, s=0$  & $\bar{c_1}=(1,0)$ &
 $e_{QC}-e_{CC}$ & + or -\\

  \rowcolor[gray]{.9} & $\bar{c_2}=(0,1)$ &
 $e_{SC} - e_{CC}$ & + or - \\

$w_{QC}$  & $\bar{u}_1$ &$e_{CQ} (e_{CC}-e_{QC})/(e_{CQ}+e_{QC}-e_{CC})
$ & - \\

  &$\bar{u}_2$ & $\xi/(e_{CQ}+e_{QC}-e_{CC}) $  & + if  $\xi>0$ or  - if $\xi<0$\\

\rowcolor[gray]{.9}  $w_{SC}$  & $\bar{v}_1$ &$(e_{CC}-e_{SC})(e_{CS}-e_{SS})/(e_{SC}-e_{CC} +e_{CS}-e_{SS} )$ & -
 \\

 \rowcolor[gray]{.9}   & $\bar{v}_2$ &$\nu/(e_{SC}-e_{CC} +e_{CS}-e_{SS})$ & + if $\nu>0$ or - if $\nu<0$\\

    \bottomrule
    \end{tabular}
\end{adjustbox}
\caption{ This table lists the eigenvectors and eigenvalues of the replicator equation for the model with three strategies computed in each equilibrium (except the coexistence equilibrium). The last column shows the possible sign that each eigenvalue could have.}

\label{corners}
\end{table}

We have that $\epsilon $ is always positive. We also have that $\nu>0$ is equivalent to:
\begin{equation}
e_{CC} > f_1(\theta,e_{SS}) := \frac{e_{QS}e_{SC} + e_{QC}e_{CS} - e_{QC}e_{SS}-e_{SC}e_{CS}}{e_{SS} - e_{QS}}
\label{f1}
\end{equation}
Similarly, $\xi>0$ is equivalent to:

\begin{equation}
e_{CC} < f_2(\theta, e_{SS}) :=  \frac{e_{SQ}e_{QC}+e_{CQ}e_{SC}-e_{CQ}e_{QC}}{e_{SQ}}
\label{f2}
\end{equation}

\begin{figure}[H]
\begin{tabular}{cc}
\includegraphics[scale=0.4]{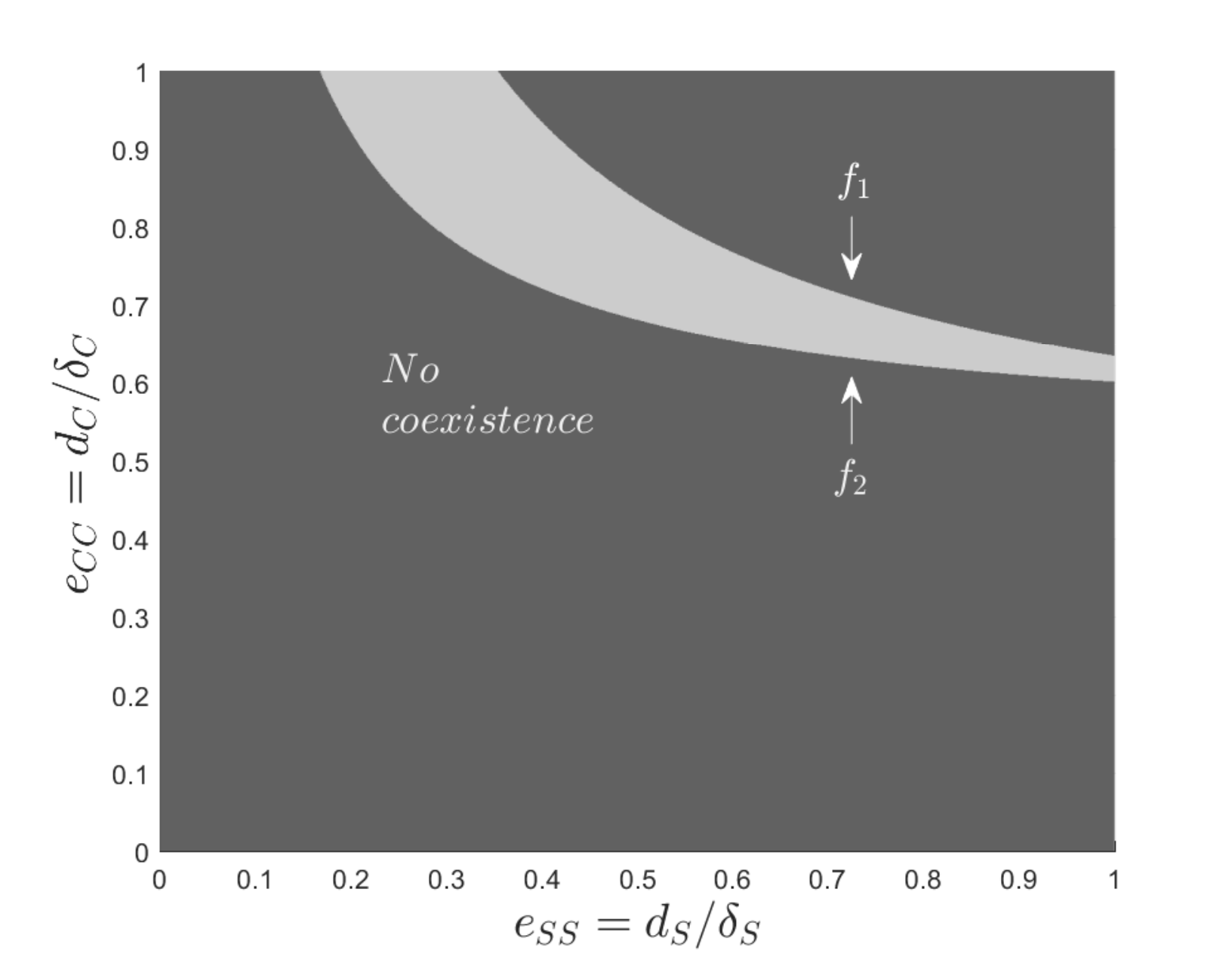}&
\includegraphics[scale=0.4]{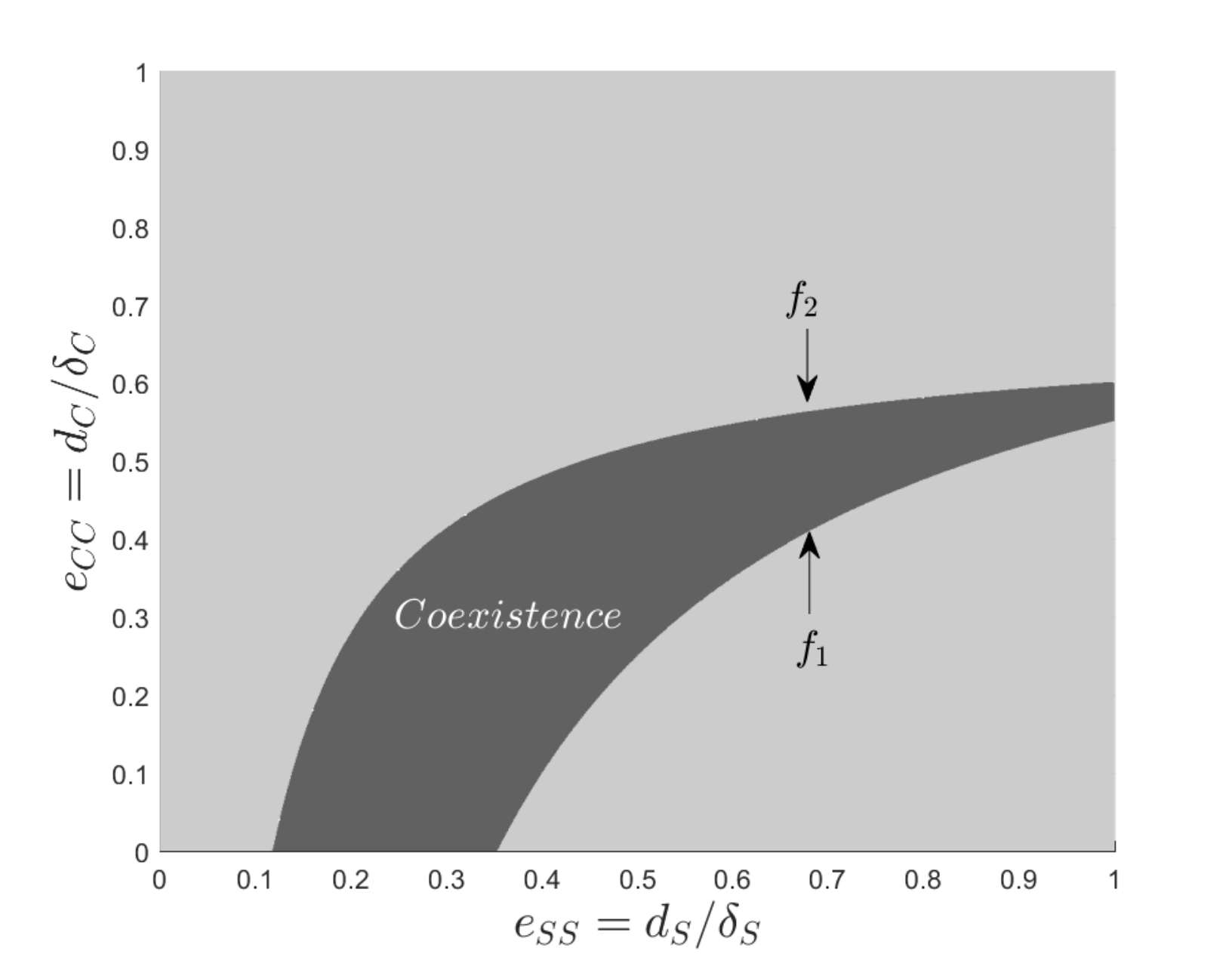}\\
$(a)$ $\theta < e_{SC}/R = 3/4$ &$(b)$ $\theta > e_{SC}/R=3/4$ \\
\end{tabular}
\caption{Possible scenarios for $f_1$ and $f_2$ depending on the value of $\theta$.}
\label{coexistence3}

\end{figure}

Figure \ref{coexistence3} shows the possible scenarios for $f_1$ and $f_2$ depending on the value of $\theta$. If $\theta < e_{SC}/R$, we have that $f_1 > f_2$ (left side in Figure \ref{coexistence3}). Besides, $f_1$ and $f_2$ are decreasing and convex. We also have that $f_1$ and $f_2$ are above $f(\theta,1)=e_{SC}$. In consequence, if $e_{CC}<e_{SC}$ (which is equivalent to the existence of $w_{SC}$), we have that $\nu<0$ and $\xi>0$. This  implies that if $\theta < e_{SC}/R$, then $w_{SC}$  is stable, $w_{QS}$ is unstable (if exists) and $w^*$ does not exist.

On the other hand, $f_1<f_2$ if $\theta > e_{SC}/R$ (right side in Figure \ref{coexistence3}). Moreover, $f_1$ and $f_2$ are increasing and concave functions that are below $f(\theta,1)=e_{SC}$. If we assume that $w_{SC}$ always exists (which is equivalent to $e_{CC}<e_{SC}$), then we have $e_{QC} = R \theta > e_{CC}$. This last inequality is equivalent to the existence of $w_{QC}$. Furthermore, if $e_{CC}<f_1$ ($\nu<0$ and $\xi>0$), the equilibrium $w_{SC}$ is stable and if $e_{CC}>f_2$ ($\nu>0$ and $\xi<0$), the equilibrium $w_{QC}$ is stable. We also have that $w^*$ exists if $f_1<e_{CC}<f_2$ (which is equivalent to $\nu>0$ and $\xi>0$). 

In the model with two strategies we have that $\hat{s}>1/2$ if $e_{CC} \leq e_{SS} - (e_{CS}-e_{SC})$. We also have that the segment $l(e_{SS}) := e_{SS} - (e_{CS}-e_{SC})$, $e_{CC} \geq 0, e_{SS} \leq 1$ is below the concave function $f_2$ for $\theta > e_{SC}/R$. Indeed, $l(1)= 1-(e_{CS}-e_{SC}) \leq e_{SC}=f_2(\theta,1)$ and $f_2(e_{CS}-e_{SC})\geq 0$. In consequence, if $\hat{s}>1/2$ for the model with two strategies, we have $e_{CC}< f_2$ and $w_{QC}$ is unstable. 

\begin{thebibliography}{99}
\bibitem{akre2011signal}  Karin L Akre, Hamilton E Farris, Amanda M Lea, Rachel A Page, and Michael J Ryan. Signal perception in frogs and bats and the evolution of mating signals. science, 333(6043):751?752, 2011.
\bibitem{akre2011female} Karin L Akre and Michael J Ryan. Female túngara frogs elicit more complex mating signals from males. Behavioral Ecology, page arr065, 2011.
\bibitem{andersson1994sexual} Malte B Andersson. Sexual selection. Princeton University Press, 1994. 300 
\bibitem{baugh2011relative} Alexander T Baugh and Michael J Ryan. The relative value of call embellishment in túngara frogs. Behavioral ecology and sociobiology, 65(2):359?367, 2011.
\bibitem{halfwerk2014risky} W Halfwerk, PL Jones, RC Taylor, MJ Ryan, and RA Page. Risky ripples allow bats and frogs to eavesdrop on a multisensory sexual display. Science, 343(6169):413?416, 2014.
\bibitem{page2008effect} Rachel A Page and Michael J Ryan. The effect of signal complexity on localization performance in bats that localize frog calls. Animal Behaviour, 76(3):761?769, 2008.
\bibitem{bernal2007cues} Ximena E Bernal, Rachel A Page, A Stanley Rand, and Michael J Ryan. Cues for eavesdroppers: do frog calls indicate prey density and quality? The American Naturalist, 169(3):409?415, 2007.
\bibitem{lea2015irrationality} Amanda M Lea and Michael J Ryan. Irrationality in mate choice revealed by túngara frogs. Science, 349(6251):964?966, 2015.
\bibitem{ryan1980female}Michael J Ryan. Female mate choice in a neotropical frog. Science, 209(4455):523?525, 1980.
\bibitem{ryan1985tungara}Michael J Ryan. The túngara frog: a study in sexual selection and communication. University of Chicago Press, 1985.
\bibitem{ryan2011replication}Michael J Ryan. Replication in field biology: the case of the frog-eating bat. Science, 334(6060):1229?1230, 2011.
\bibitem{tarano2015choosing} Zaida Tárano. Choosing a mate in a cocktail party-like situation: The effect of call complexity and call timing between two rival males on female mating preferences in the túngara frog physalaemus pustulosus. Ethology, 121(8):749?759, 2015.
\bibitem{weibull1997evolutionary}J\"{o}rgen W Weibull. Evolutionary game theory. MIT press, 1997.
\end{thebibliography}
\end{document}